\documentclass[twocolumn,prl,tightenlines,nofootinbib,superscriptaddress,showpacs]{revtex4}

\usepackage{amsmath}
\usepackage{amssymb,amsfonts,latexsym}
\usepackage{bm}
\usepackage[mathcal]{euscript}
\usepackage{graphicx}
\usepackage{epsfig}
\usepackage{color}

\begin{document}

\title{Collective Motion of Vibrated Polar Disks}

\author{Julien Deseigne}
\affiliation{Service de Physique de l'Etat Condens\'e, CEA-Saclay, URA 2464 CNRS, 91191 Gif-sur-Yvette, France}
\author{Olivier Dauchot}
\affiliation{Service de Physique de l'Etat Condens\'e, CEA-Saclay, URA 2464 CNRS, 91191 Gif-sur-Yvette, France}
\author{Hugues Chat\'{e}}
\affiliation{Service de Physique de l'Etat Condens\'e, CEA-Saclay, URA 2464 CNRS, 91191 Gif-sur-Yvette, France}
 
\date{\today} 

\begin{abstract}
We experimentally study a monolayer of vibrated {\it disks} with
a built-in polar asymmetry which enables them to move quasi-balistically
on a large persistence length. Alignment occurs during collisions as a result of 
self-propulsion and hard core repulsion. Varying the amplitude of the vibration,
we observe the onset of large-scale collective motion and the existence
of giant number fluctuations with a scaling exponent in agreement with
the predicted theoretical value.
\end{abstract}

\maketitle

The recent surge of theoretical/numerical activity about 
the collective properties of interacting self-propelled particles
has produced some striking results, even in the simplest
situations where local alignment, the only interaction, is 
competing with some noise: for instance, true long-range
order may arise in two dimensions, yielding collectively-moving
ordered phases endowed with generic long-range correlations and
anomalous ``giant'' number fluctuations \cite{TTR-review,GNF,VICSEK,CHATE}.
Despite the ubiquity of the collective motion, observed at 
all scales in more or less complex situations ranging from
the cooperative action of molecular motors \cite{MOTORS},
the collective displacement of cells \cite{CELLS}, 
to the behavior of large animal, human, or robot groups \cite{GROUPS}, 
there is, as of now, a lack of well-controlled experiments to which this
theoretical progress can be seriously confronted. 
Indeed, working with large animal groups usually implies that experiments are
just observations, with the unavoidable difficulties to track trajectories
and without much of a control parameter to vary \cite{STARFLAG}. 
Similarly, experiments on living cells (during development or wound healing or within 
bacteria and amoeba colonies) often involve the presence of external (chemical) gradients, 
genetic factors, etc., which are hard to evaluate and known to have a possibly strong
influence. There is hence a crucial need for {\it model experiments} using 
man-made objects with rather well understood interactions.
Swimmers \cite{SWIMMERS} (from chemically-powered nanorods to 
microscopic and macroscopic size mechanical devices) offer an 
interesting direction but the intrinsically long-range nature of
hydrodynamic interactions may appear as a unnecessary complication. 
In this context, vibrated, dry, inert, ``granular'' particles appear
as an attractive case where much control can be exerted on the system, in the
absence of long-range interactions or unwanted additional features, so that
the onset of collective motion would then be a bona fide spontaneous symmetry 
breaking phenomenon.

Various objects can be set in fairly regular motion on a flat surface
when vibrated properly:
Yamada, Hondou, and Sano were pioneers in demonstrating that an axisymmetric polar
object vibrated between two plates can move quasi-ballistically
\cite{SANO}. 
%Later, Dorbolo {\it et al.} showed that an axisymmetric {\it apolar}
%object (a dimer) can be brought into a similar drift mode \cite{DORBOLO}.
At the collective level, Kudrolli's group studied the behavior of polar rods
\cite{KUDRO2008} 
and, more recently, of short snake-like chains \cite{KUDRO2010}, 
but was unable to observe genuine long-range orientational order, 
{\it i.e.} collective motion. A few other works have dealt with 
the collective properties of shaken elongated apolar particles 
(a realization of so-called ``active nematics'')\cite{EXP-RODS}, 
but there no net collective motion is expected anyway.
%%%XXX check validity of statement wrt rod "vortices" (Kudro?)

Thus, to our knowledge, no well-controlled experiment has produced 
a fluctuating, collectively-moving ordered phase of the type frequently observed in simple 
numerical models. This may be just due to the scarcity of attempts, but recent
results might provide a deeper reason: it was found that 
self-propelled particles with apolar (nematic) alignment 
interactions cannot give rise to polar order, i.e. to collective motion \cite{THEO-RODS}. 
(They may give rise, however, to nematic order.) The few experiments mentioned above
all dealt with elongated objects (``self-propelled rods''), and fall into
this class because of their shape.

\begin{figure}[t!]
\includegraphics[width=0.9\columnwidth]{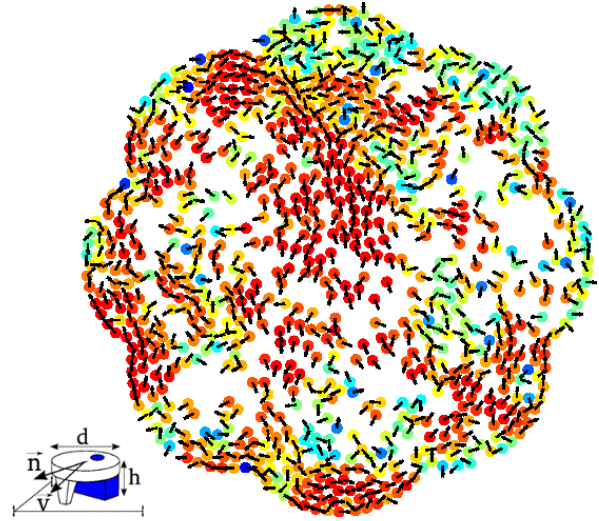}
\caption{Collective motion of self-propelled disks (color online).
Bottom left: sketch of our polar particles. Main panel: 
snapshot of an ordered regime observed in our flower-shape domain.
The color scale reflects the local polar order from perfect alignment (red) to
anti-alignment (blue). The intrinsic polarity of the particles is
indicated by the black arrows.}
\label{fig1}
\vspace{-5mm}
\end{figure}

In this Letter, we report on experiments conducted on vibrated {\it disks} with
a built-in polar asymmetry which enables them to move coherently (Fig.~\ref{fig1}).
The isotropic shape of the particles and their rather specific inelastic collision
properties prevent strong nematic alignment.
Varying the amplitude of the vibration, we observe the onset of large-scale 
collective motion and the existence of giant number fluctuations 
with a scaling exponent in agreement with the predicted theoretical value. 
We discuss the difficulties in characterizing collective motion
in a finite domain and the possible key differences with the simple
models usually considered at the theoretical level.

Experiments with shaken granular particles are notoriously
susceptible to systematic deviations from pure vertical vibration~\cite{IGOR}.
We use a $110$ mm thick truncated cone of expanded polystyren sandwiched 
between two nylon disks. The top disk (diameter $425$ mm) is 
covered by a glass plate on which lay the particles. The bottom one
(diameter $100$ mm) is mounted on the slider of a stiff square
air-bearing (C40-03100-100254,IBSPE), which provides 
virtually friction-free vertical motion
and submicron amplitude residual horizontal motion. 
The vertical alignment is controlled by set screws. 
The vibration is produced with an electromagnetic servo-controlled shaker 
(V455/6-PA1000L,LDS), the accelerometer for the control being fixed at
the bottom of the top vibrating disk, embedded in the expanded polystyren. 
A $400$ mm long brass rod couples the air-bearing slider and the shaker.
It is flexible enough to compensate for the alignment mismatch, but stiff
enough to ensure mechanical coupling.
The shaker rests on a thick wooden plate ballasted with $460$ kg of lead bricks
and isolated from the ground by rubber mats (MUSTshock 100x100xEP5,Musthane). 
We have measured the mechanical response of the whole setup and found no resonances in the
window $70-130$ Hz. Here, we use a sinusoidal vibration of frequency $f=115$ Hz and
vary the relative acceleration to gravity $\Gamma = 2\pi a f^2/g$. The
vibration amplitude $a$ at a peak acceleration of 1 $g$ at this frequency is $25$
$\mu$m. Using a triaxial accelerometer (356B18,PCB Electronics), 
we checked that the horizontal to vertical ratio is lower
than $10^{-2}$ and that the spatial homogeneity
of the vibration is better than $1\%$.

Our polar particles are micro-machined copper-beryllium disks
(diameter $d = 4$ mm) with an off-center tip and a glued rubber skate 
located at diametrically opposite positions (Fig.~\ref{fig1}). 
These two "legs", which have different mechanical response under 
vibration, endow the particles with a polar axis which can be 
determined from above thanks to a black spot located on their top.
Under proper vibration, they can be set in directed motion (see below).
Of total height $h=2.0$ mm, they 
are sandwiched between two thick glass plates separated by a gap of $H=2.4$ mm. 
We also used, to perform ``null case experiments'', 
plain rotationally-invariant disks (same metal, diameter, 
and height), hereafter called the ``symmetric'' particles.
We laterally confined the particles in a flower-shaped arena of 
internal diameter $D=160$ mm (Fig.~\ref{fig1}). The petals
avoid the stagnation and accumulation of particles along the boundaries as 
reported for instance in~\cite{KUDRO2008} by ``reinjecting'' 
them into the bulk.
A CCD camera with a spatial resolution of 1728 x 1728 pixels and standard
tracking software is used to capture the motion of the particles
at a frame rate of $20$ Hz.
% This allows us to
% measure both the position $\vec{r}_i(t)$ and polarity
% $\vec{n}_i(t)$ of the particles (thanks to the off-center black dot painted
% on their top).
In the following, the unit of time is set to be the period of 
vibration and the unit length is the particle diameter.
Within these units, the resolution on the position $\vec{r}$ of the
particles is better than $0.1$, that on the orientation $\vec{n}$
is of the order of $0.05$ rad and the lag separating two images is $\tau_0=5.75$. 
Measuring the long-time averaged spatial density map (for various numbers of particles),
we find that this density field slightly increases near the boundaries, but
is constant to a few percent in a region of interest (ROI) of diameter $20d$.
This provides an additional check of the spatial homogeneity of our setup.

\begin{figure}[t] 
\begin{center} 
\vbox{
\hbox{\hspace{1.95cm}(a)\hspace{4.0cm} (b)} 
\includegraphics[clip,width=8.4cm]{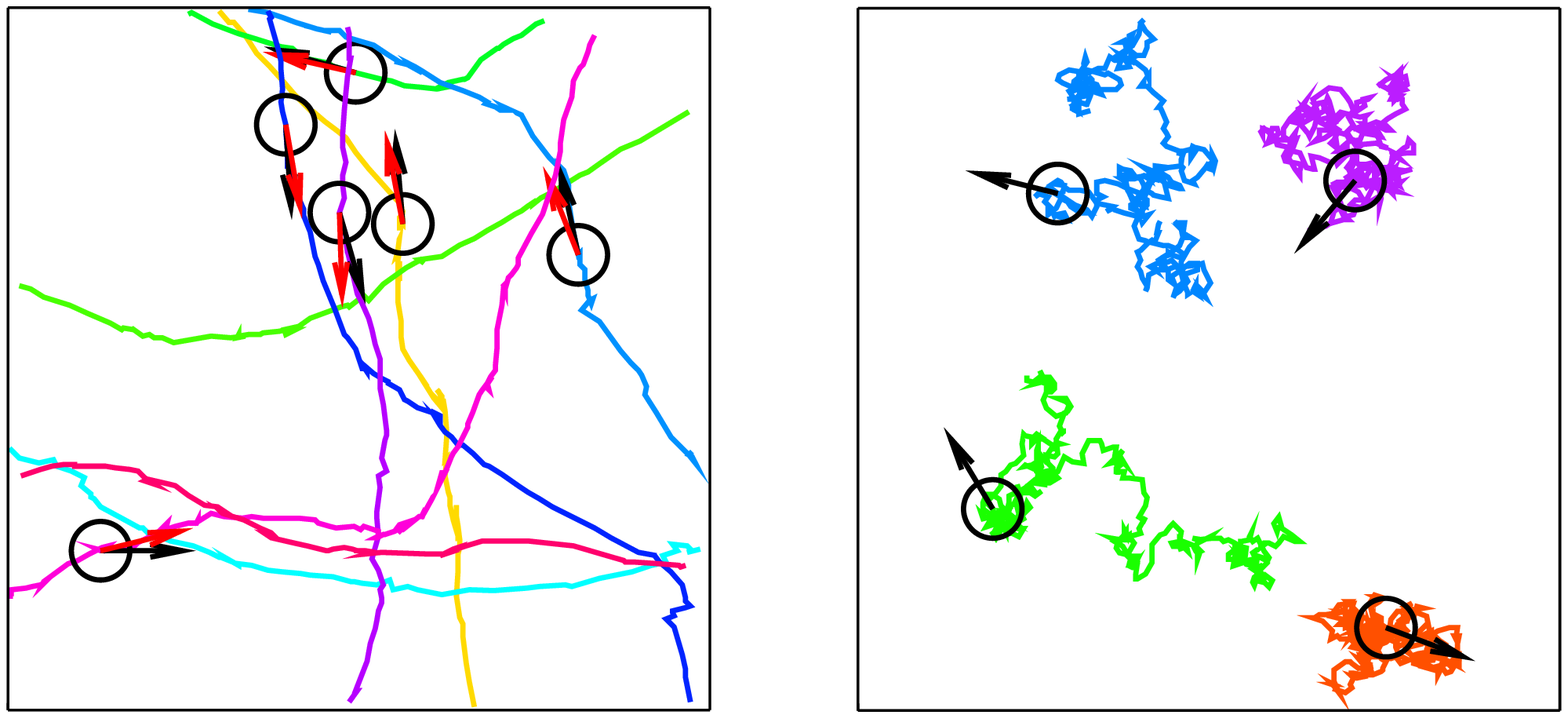}
}
\includegraphics[clip,width=8.4cm]{Fig2-new.eps}
\caption{(color online) Individual dynamics for $\Gamma=2.7$.
(a) typical portions of polar particles trajectories inside the ROI. 
Black and red arrows indicate  $\vec{v}_i^t$ and  $\vec{n}_i^t$ 
at selected times. The domain area is about $15\times 15 d$.
(b) same for symmetric particles. 
(c) pdf (lin-lin) of $\alpha$, the angle between $\vec{v}_i^t$ and  $\vec{n}_i^t$.
(d) variation of angular diffusion coefficient $D_\theta$ with $\Gamma$.
}
\label{fig2} 
\end{center} 
\vspace{-8mm}
\end{figure} 

We first performed experiments with $50$ particles, i.e. at a
surface fraction small enough so that collisions are rare and the individual 
dynamics can be investigated. For large acceleration, the polar particles
describe random-walk like trajectories with short persistence length. 
Decreasing $\Gamma$, they show more and more directed motion,
and the persistence length quickly exceeds the system size. 
This is in contrast with the  symmetric particles which retain 
the same shortly correlated individual walk dynamics for all $\Gamma$ values
(Fig.~\ref{fig2}ab).

More precisely, individual velocities
$\vec{v}_i(t) \equiv (\vec{r}_i(t+\tau_0)-\vec{r}_i(t))/\tau_0$
measured within the ROI have a well-defined
most probable or mean value $v_{\rm typ}\simeq 0.025$ 
which changes by only 6\% over the interval $\Gamma\in [2.7, 3.7]$ (not shown). 
For smaller values of $\Gamma$ the velocity decreases suddenly 
and the particles come to an almost complete stop around $\Gamma=2.4$.
The local displacements of our polar particles are overwhelmingly taking place along
$\vec{n}_i(t)$, their instantaneous polarity (Fig.~\ref{fig2}c).
The distribution of the angle $\theta_i(t,t+\tau_0)$ by which they turn
during an interval $\tau_0$ (defined using the polarity $\vec{n}_i(t)$)
is an exponential distribution of zero-mean and variance $2D_\theta/\tau_0$.
The angular diffusion constant $D_\theta$ decreases fast and linearly 
for $\Gamma\in [2.7, 3.7]$ (Fig.~\ref{fig2}d). 
In contrast again, $D_\theta$ is about one order of magnitude larger for our 
isotropic particles, and varies little with $\Gamma$ (not shown).
A persistence length can then be defined as
$\xi = \frac{1}{2}\pi^2 v_{\rm typ}/D_\theta$
(i.e. the length traveled over the time needed to turn by $\pi$, assuming a 
constant speed $v_{\rm typ}$). Its typical value decreases from above $100$
for $\Gamma=2.7$ to around $20$ for $\Gamma=3.7$ whereas it stays around 1
for the symmetric particles.

\begin{figure}[t] 
\includegraphics[clip,width=4cm]{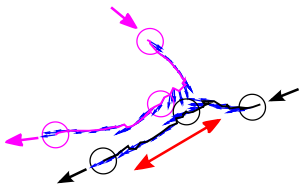}
\vspace{-0.5cm}
\caption{(color online) Trajectories of two particles ``during'' a collision: 
they first collide almost head-on, but repeated contacts 
(all along the red arrow) finally leave them almost aligned, 
despite their isotropic shape.}
\label{fig3} 
\end{figure}

We now turn to the collective dynamics of our polar particles.
As seen above, the relative acceleration $\Gamma$ has a strong 
influence on their individual
dynamics, controlling the persistence length of their trajectories
via the angular diffusion constant $D_\theta$.
During collisions, they typically bounce against each other several times,
yielding, on average, some degree of alignment (Fig.~\ref{fig3}).
All this is reminiscent of Vicsek-like models, for which 
one of the main control parameters is the strength of the angular 
noise competing with the alignment interaction \cite{VICSEK,CHATE}. 
Thus $\Gamma$ is not only an easy control parameter,
but also a natural one, which we use in the following. 
The surface fraction $\phi$ of particles is another natural control parameter
in collective motion and granular media studies, but it is somewhat
more tedious to vary, and, more importantly, one should avoid to
deal with too few, respectively too many, particles in 
order to prevent loss of statistical quality, respectively jamming effects. 
Below, we present results obtained with $N=890$ particles, which 
gives a surface fraction  $\phi\simeq 0.38$ in the ROI where
an average of 160 particles (slightly dependent on $\Gamma$) 
is found. Similar results were obtained at nearby densities.
To characterize orientational order, we use the modulus of
the average velocity-defined polarity
$\Psi(t) =| \langle \vec{u}_i(t) \rangle |$
where $\vec{u}_i(t)$ is the unit vector along  $\vec{v}_i(t)$ and 
the average is over all particles inside the ROI at time $t$. \cite{NOTE}

\begin{figure}[t] 
\begin{center} 
\includegraphics[clip,width=8.6cm]{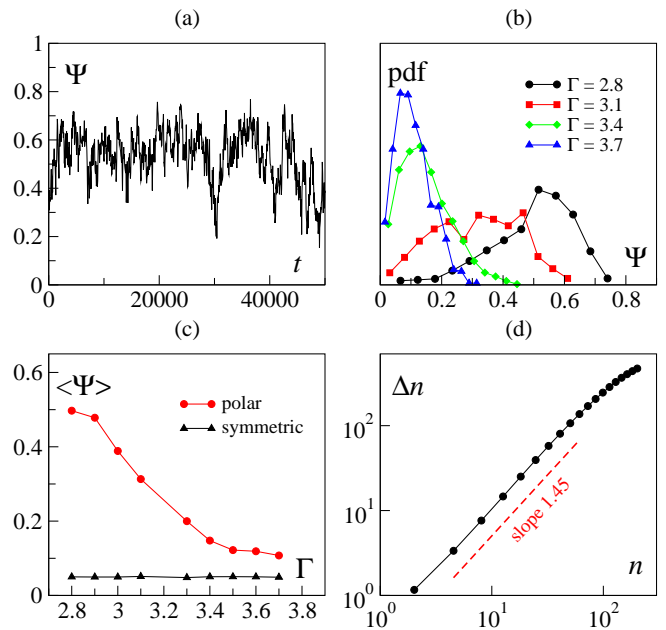} 
\caption{(color online) Collective dynamics.
(a) time series of order parameter $\Psi$  at $\Gamma=2.8$.
(b) pdf (lin-lin) of $\Psi(t)$ at various $\Gamma$ values.
(c) $\langle\Psi\rangle$ vs  $\Gamma$ for polar and symmetric
 particles.
(d) $\Delta n$ vs $n$  at $\Gamma=2.8$.}
\label{fig4}
\end{center} 
\vspace{-8mm}
\end{figure}

At low $\Gamma$ values, for which the directed motion of our polar 
particles is most persistent, we observe spectacular large-scale
collective motion, with jets and swirls as large as the 
system size (Fig.~\ref{fig1} and \cite{EPAPS}). 
Of course, because our boundary conditions are not periodic,
the collective motion observed is not sustained at all times.
Large moving clusters form, then breakdown, etc. 
As a result, the times series of the order parameter $\Psi$ 
presents strong variations, but can take a rather
well-defined order one value for long periods of time 
(Fig.~\ref{fig4}a). At high $\Gamma$ values (large noise)
no large-scale ordering is found. Decreasing
$\Gamma$, the pdf of $\Psi$ becomes wider and wider,
with a mean and a most probable value increasing sharply (Fig.~\ref{fig4}bc).
Note that the most probable value corresponds, at small $\Gamma$,
to the plateau value found in time series of  $\Psi$.
%%%XXX show both mean and mpv on graph??
%%%XXX HERE PUT BACK CORREL FCTS IF SPACE ALLOWS
Thus, we observe the clear emergence of long-range orientational 
order over the range of usable $\Gamma$ values.
In contrast, the same experiments realized with our symmetric 
particles do not give rise to any collective motion
(Fig.~\ref{fig4}c), which ultimately indicates that our observations 
with polar particles are not due to some residual
large-scale component of our shaking apparatus \cite{IGOR}.

Unfortunately, we could not observe the saturation of the
order parameter expected deep in the ordered phase, because the
``self-propulsion'' of our polar particles deteriorates 
for $\Gamma\lesssim 2.7$.
Nevertheless, large  $\Psi$ values were observed, signalling
that our lowest usable $\Gamma$ values are already in the
ordered phase, albeit not quite surely out of the critical/transitional
region. We thus investigate the emergence of the so-called
``giant number fluctuations'' (GNF) which have been shown
theoretically and numerically to be a landmark of orientationally-ordered 
phases for active particles \cite{TTR-review,GNF}. To this aim, we
recorded, along time, the number $n(t)$ of particles present 
in boxes of various sizes located within the ROI. 
GNF are characterized by the fact that the variance $\Delta n$ 
of this number scales faster than the mean $n$. 
This is indeed what we find: over a range of scales, $\Delta n$ 
grows like $n^\alpha$ with $\alpha\sim1.45\pm 0.05$.
For larger scales, one feels the finite system size and $\Delta n$ levels off.
In fact, according to the prediction derived from the work of 
Tu and Toner \cite{TCGR} and confirmed in simulations \cite{CHATE},
this number should be $1.6$. Thus our finding is quite consistent
with the predicted value, all the more so since $\alpha$ is expected
to converge from below as the system size increases \cite{GNF}.
Although this will require to be confirmed by experiments 
performed in larger dishes, this result constitutes the first
experimental evidence for GNF in collections of polar 
active particles \cite{NOTE2}. 

To summarize, we have shown that shaken particles with a polarity not
related to their shape can exhibit collective motion on scales or the order
of the domain in which they evolve. In the most ordered regimes reachable, 
we recorded giant number fluctuations with a scaling exponent 
consistent with that of polar active phases. 

That we observe dominant {\it polar} order is worth discussing: 
it was recently shown that if their alignment interaction is {\it nematic},
polar particles cannot order polarly and only nematic order arises, even
if the particles are pointwise \cite{THEO-RODS}. This nematic order
is made of polar packets \cite{PERUANI}, which could dominate the global order
in a small domain such as our ROI. It is not clear, at this point,
whether our system falls in this class. 
As a matter of fact, we do observe a small fraction
of particles going ``against'' the main flow in our most ordered regimes
(Fig.~\ref{fig1} and \cite{EPAPS}). But this remains rare ---most of the time,
polar alignment is observed, see Fig.~\ref{fig3}---, in contrast
to the above-mentioned studies, but in line with the numerical work of
Grossman, Aranson, and Ben Jacob \cite{GROSSMAN}.
 Further investigations will require 
a detailed study of the statistics of collisions.

In Vicsek-style models (and their continuous descriptions) 
no GNF proper exist near the transition, where
high-order high-density bands emerge \cite{CHATE,BGD}.
Here, we found GNF in our most-ordered regimes, with approximately
the expected exponent. But it is impossible, at this stage, to disentangle
fluctuations due to the proximity of the transition, those due to the
frustration induced by our boundaries (which would break bands),
and ``genuine'' GNF.  Thus, performing experiments in larger
domains and looking for parameter values which would allow to
go deep into the ordered phase is of utmost importance.
This could also allow to study the nature of the transition to
collective motion. To this aim, an even better control of
our vibration table is necessary, a task we are currently
pursuing. 

We thank M.~Van Hecke for advice in the design of our system,
V.~Padilla and C.~Gasquet for technical assistance, and 
E.~Bertin for enlightening discussions. 
Work supported by the French ANR project DyCoAct.


\begin{thebibliography}{99}
\bibitem{NOTE} Using $\vec{u}$ and not $\vec{n}$ allows to 
calculate the same quantity for both polar and 
symmetric particles. For the polar particles,
using $\vec{n}$ yields similar results.
\end{thebibliography}
\end{document}